\documentclass[aps,prl,twocolumn,amsfonts,amsmath,floatfix,showpacs,superscriptaddress,footnoteintext]{revtex4}

\usepackage[final]{graphicx}
\usepackage{color}

\newcommand{\be}{\begin{equation}} \newcommand{\ee}{\end{equation}}

\begin{document}

\title{Conformal Field Theory of Critical Casimir Interactions in 2D}

\date{\today}

\author{Giuseppe Bimonte}
\affiliation{Dipartimento di Scienze
Fisiche, Universit{\`a} di Napoli Federico II, Complesso Universitario
MSA, Via Cintia, I-80126 Napoli, Italy}
\affiliation{INFN Sezione di
Napoli, I-80126 Napoli, Italy }

\author{Thorsten Emig}
\affiliation{Laboratoire de Physique
Th\'eorique et Mod\`eles Statistiques, CNRS UMR 8626, B\^at.~100,
Universit\'e Paris-Sud, 91405 Orsay cedex, France}

\author{Mehran Kardar} \affiliation{Massachusetts Institute of
Technology, Department of Physics, Cambridge, Massachusetts 02139, USA}

\begin{abstract}
  Thermal fluctuations of a critical system induce long-ranged Casimir
  forces between objects that couple to the underlying field.  For two
  dimensional (2D) conformal field theories (CFT) we derive an exact
  result for the Casimir interaction between two objects of arbitrary
  shape, in terms of (1) the free energy of a circular ring whose
  radii are determined by the mutual capacitance of two conductors
  with the objects' shape; and (2) a purely geometric energy that is
  proportional to conformal charge of the CFT, but otherwise
  super-universal in that it depends only on the shapes and is
  independent of boundary conditions and other details.
\end{abstract}


\maketitle

Objects embedded in a medium constrain its natural fluctuations, resulting in fluctuation-induced forces~\cite{Kardar:1999a}. 
The most naturally occurring examples result from modification of electromagnetic fluctuations, manifested variously in van der Waals interactions~\cite{Parsegian:2005ly} (between atoms and molecules) to Casimir forces (between conducting plates)~\cite{Bordag:2009ve}. While fluctuations of the latter are primarily quantum in origin, thermal fluctuations of correlated fluids lead to similar interactions, most notably at a critical point (where correlation lengths are macroscopic)~\cite{Gennes:1978qf,Krech:1994kl}. 
Critical fluctuation-induced forces have been observed in helium~\cite{Garcia:1999}
and in binary liquid mixtures~\cite{Law:1999,Pershan:2005,Hertlein:2008}
Critical fluctuations of a binary mixture were recently employed to manipulate and assemble colloidal particles~\cite{Soyka:2008}.

Biological membranes are mainly composed of mixtures of lipid molecules,
and could potentially be poised close to a critical point demixing 
point~\cite{Baumgart:2007,Veatch:2007}, in the two-dimensional Ising universality class.
It has been suggested that membrane concentration fluctuations could thus
lead to critical Casimir forces between inclusions on such membranes,
motivating computation of such forces between discs
embedded in the critical Ising model~\cite{Machta:2012fu}.
Membranes (and interfaces) also undergo thermal shape fluctuations 
governed by the energy costs of bending (and surface tension)~\cite{Piran:1989}.
Modification of these fluctuations have also been proposed as
a source of interactions amongst inclusions on membranes~\cite{Goulian:1993,Golestanian:1996}, possibly accounting for patterns of colloidal particles
at an interface~\cite{Bresme:2007fk}. There is extensive literature on this topic, and the interested
reader can consult recent publications~\cite{Deserno:2012,Noruzifar:2013}.
Yet another entropic force is proposed to act between surface/membrane bio-adhesion bonds~\cite{Farago:2011}.

Conformal field theories (CFTs) have proved highly successful in studies of two dimensional (2D)
systems at criticality~\cite{Friedan:1984,Cardy:1989}.
Various boundary conditions have been examined for Ising (or 3-state
Potts) model on a cylinder~\cite{Cardy:1986fu}. Connections to Casimir forces
between parallel plates ~\cite{Kleban:1991ff,Kleban:1996pi} and spheres \cite{Burkhardt:1995cr,Eisenriegler:1995nx}
have been explored.
Non-spherical particles at large separations have been studied with
the small
particle operator expansion \cite{Eisenriegler:2004fk,Eisenriegler:2006uq}.
However, a general formulation for interactions between two (or more) objects
of arbitrary shape embedded in a CFT appears to be lacking. 
Some special cases recently studied include
interactions between two spherical holes in a free field~\cite{Rothstein:2012mi},
between circular inclusions ~\cite{Machta:2012fu} 
and needles~\cite{Vasilyev:2012dz} in a critical Ising system.
(We note in passing exact solutions for Casimir interactions between spheres in three dimensions~\cite{Burkhardt:1995cr,Eisenriegler:1995nx,Bimonte:2012hc}.)
Starting with the solution of the Laplace equation with two inclusions of
arbitrary shape as equipotentials, the system can be conformally mapped
either to a cylinder, or an annulus. We demonstrate that such mapping can
be employed to compute the Casimir interaction between the two objects
embedded in any CFT.

We consider a general two dimensional classical field theory with an
energy that is invariant under conformal transformations. Examples
include free theories, such as the capillary-wave Hamiltonian that
describes deformations with small gradients around a flat interface,
and interacting theories, like the Ising model at its critical point.
The corresponding CFT is assumed to couple to two compact objects covering
areas $S_1$ and $S_2$ via conformally invariant boundary conditions on
the boundaries $\partial S_\alpha$ ($\alpha=1$ or $2$). 
Examples include Dirichlet or Neumann conditions for a free field,
and pinned or free conditions for the Ising model. In the
following we assume that the boundaries $\partial S_\alpha$ are Jordan
curves \footnote{A Jordan curve is any non-intersecting closed planar trajectory.}.

\begin{figure}[h]
\includegraphics[width= .95\columnwidth]{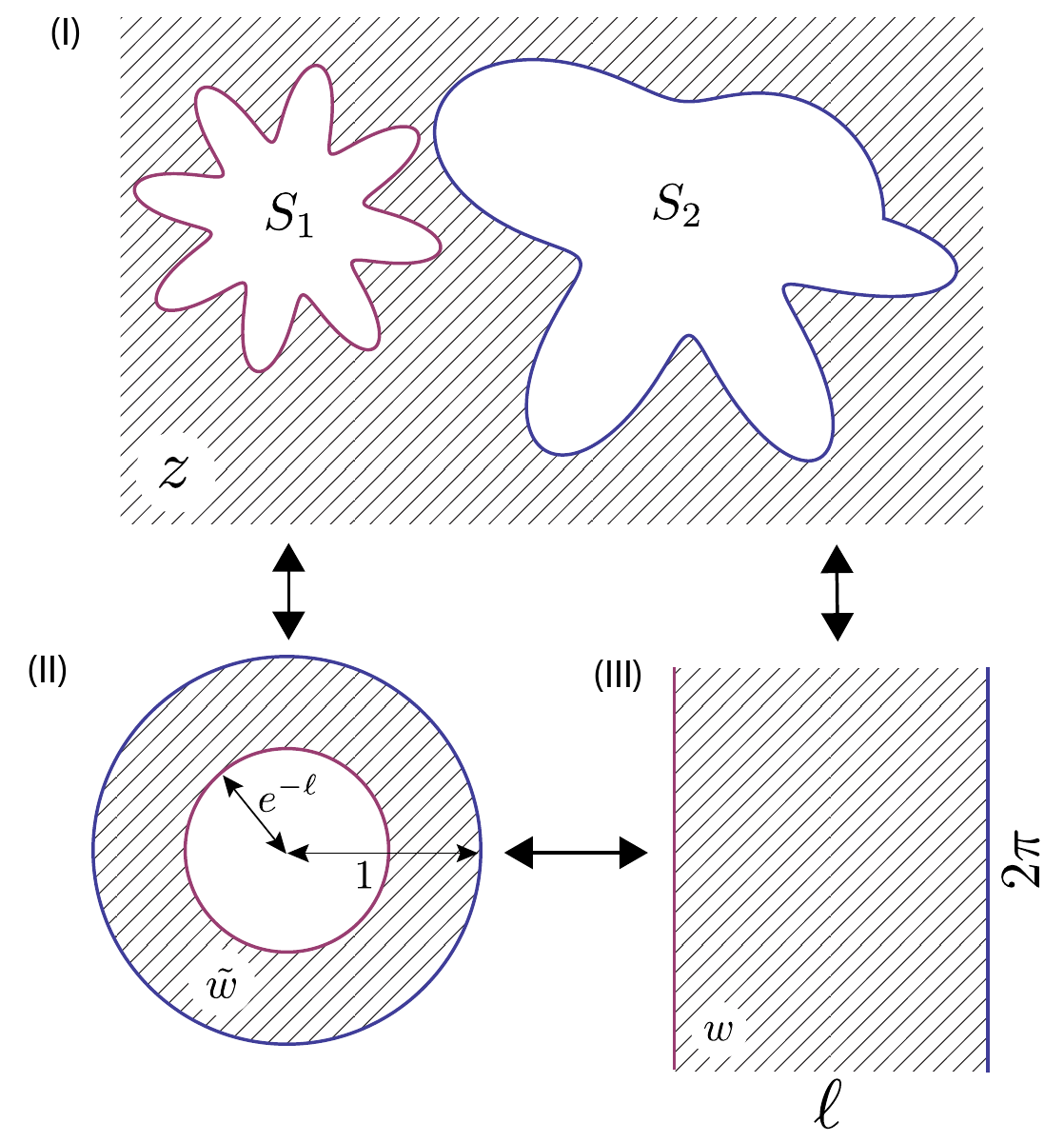}
\caption{Conformal maps of the exterior region of two objects $S_1$ and 
$S_2$ to an annulus via $\tilde w(z)$, and to the surface of a cylinder
by $w(z)$ (see text for details).}
\label{fig:conf_map}
\end{figure}

Before explaining the main steps of the derivation, and presenting
examples, we summarize our main result: The doubly connected domain
bounded by $\partial S_1$ and $\partial S_2$ can be conformally mapped
to the surface of a cylinder with unit radius and length $\ell$, or
alternatively to an annulus with outer and inner radii of $1$ and
$e^{-\ell}$, respectively, see Fig.~\ref{fig:conf_map}.  The map
$w(z)$ to the cylinder has an electrostatic interpretation: The real
(and imaginary) part of the map is $2\pi/Q$ times the
electrostatic potential (and its conjugate function) outside the
objects with the potential set to $-1$ for $\partial
S_1$, and $0$ for $\partial S_2$, with net charges of $-Q$ and
  $+Q$, respectively~\cite{Courant:1950fk}.  The cylinder length
$\ell=2\pi/C$ is then given by the mutual capacitance $C$ of two
cylindrical conducting surfaces in 3D that have the areas $S_j$ as
their cross section.  The map to the annulus is then $\tilde w(z) =
\exp[w(z)]$.  Our main result is that the $x$ and $y$ components of
the Casimir force between the two objects are combined into the
complex force
\begin{equation}
  \label{eq:1}
  F \equiv\frac{F_x-iF_y}{2}= -\partial_\zeta  {\cal F}_\text{ann.} - \frac{i c}{24\pi}
  \oint_{\partial S_2} \!\!\!\{ \tilde w,z \} dz  \,.
\end{equation}
In the first contribution above, ${\cal F}_\text{ann.}$ is the free
energy of the CFT on the annulus with the
boundary conditions of $\partial S_1$ ($\partial S_2$) on the
  inner (outer) circle (see below for examples); and the derivative
is with respect to $\zeta=(x_2-x_1)+i(y_2-y_1)$, the distance in the
complex plane between two origins $(x_\alpha,y_\alpha)$ on the
objects.  (Note that throughout the paper we set $k_BT=1$, such that
${\cal F}=-\ln Z$.)  The second term is proportional to $c$, the
conformal charge of the CFT, and involves the integral of the
Schwarzian derivative~\cite{Francesco:1997kx} of the conformal map
$\tilde w$, $\{ \tilde w ,z\} \equiv(\tilde w'''/\tilde w') - (3/2)
(\tilde w''/\tilde w')^2$, along the counter $\partial S_2$ performed
counter-clockwise.  This contribution to the force can be written in terms
of a `geometric' free energy as $F_\text{geo}=-c\partial_\zeta {\cal
  F}_\text{geo.}$.  $ {\cal F}_\text{ann.}$ varies with the CFT but
depends on geometry only through the capacitance via $\ell=2\pi/C$.
By contrast $ {\cal F}_\text{geo.}$ is fully determined by the shape
of the objects, independently of the CFT. In this sense,
$F_\text{geo}$ is {\it super-universal} as it is the same for all
CFT's (up to a factor of $c$).  It vanishes if and only if $\tilde w$
is a {\it global} conformal map, i.e., when the objects $S_\alpha$ are
circular. This follows as the Schwarzian derivative measures the
deviation of the map from being global.

{\it Sketch of proof} --- We begin by relating the change in the cylinder
length $\ell$ with the objects separation $\zeta$, to the map $w(z)$.
After a small displacement of $S_2$, the electrostatic energy is modified by
\begin{equation}
  \label{eq:3}
  \delta {\cal E_\text{el}} = \frac{1}{2\pi i} \oint_{\partial S_2}\!\!\! \alpha(z) \, T_\text{el}(z) dz
  + \text{c.c.}\,,
\end{equation}
where $\alpha(z)$ reverses the motion.
The electrostatic stress tensor is well known and can be expressed in terms
of  the cylinder map $w(z)$ by $T_\text{el}(z) = -(\pi/2) (\partial_z w)^2$. 
Since at fixed charges $Q=\pm 2\pi$, $\delta {\cal E_\text{el}}=-(2\pi)^2\delta (1/2C)$,
and $\ell=2\pi/C$, $\delta \ell = -\delta {\cal E_\text{el}} /\pi$. 
By applying Eq.~\eqref{eq:3} within
$S_2$ with $\alpha = -\delta x$ and $\alpha = - i \delta y$, and setting 
$\partial_\zeta \ell = (\partial_x \ell- i \partial_y \ell)/2$, we then find
$\partial_\zeta \ell = (i/4\pi) \oint_{\partial S_2} (\partial_z w)^2  dz$.

The displacement of  $S_2$  changes the Casimir free energy by an amount 
$\delta {\cal F}$, also given by Eq.~\eqref{eq:3} with $T_\text{el}$ replaced 
by the stress tensor $T(z)$ of the CFT outside the objects. 
To obtain a simple expression for $T(z)$ in terms of the above maps we proceed
as follows:
As in Eq.~\eqref{eq:3}, the stress tensor for the cylinder can be expressed
in terms of the derivative of the free energy with respect to its length by
$2T(w)=\partial_\ell {\cal F}_\text{cyl.}=\partial_\ell {\cal F}_\text{ann.}-c/12$.
For the second form, we have relied on a known relation 
${\cal F}_\text{cyl.}={\cal F}_\text{ann.}-\ell c/12$ between the cylinder and
annulus free energies~\cite{Cardy:1989,Francesco:1997kx}.
Next, we note that for any map $w(z)$, the stress tensor transforms
according to  $T(z)=(\partial_z w)^2 T(w) + (c/12) \{ w,z\}$~\cite{Francesco:1997kx}.
We can use this expression to relate $T(\tilde w)$ to $T(w)$, and
separately to relate $T(z)$ to $T(\tilde w)$, to
finally obtain $T(z) = (1/2) (\partial_z w)^2 \partial_\ell {\cal F}_\text{ann.} 
+ (c/12) \{ \tilde w,z\}$. Using this result in Eq.~\eqref{eq:3} both with $\alpha =
-\delta x$ and $\alpha = - i \delta y$ we arrive at
Eq.~\eqref{eq:1} after using the previous expression for
$\partial_\zeta \ell$.

{\it Asymptotic limits of the annulus free energy} ---
The scaling of  ${\cal F}_{\text{ann.}}$ for small and large
$\ell=2\pi/C$ (and hence short and large separations $|\zeta|$) can be
obtained from two equivalent representations of one-dimensional
quantum field theories (QFT's) (page~423 of Ref.~\cite{Francesco:1997kx}). 
First, consider the QFT on a circle of
circumference $\delta$ with Hamiltonian $\hat H = (2\pi/\delta) (\hat
L_0 + \hat{\bar L}_0- c/12)$ where $\hat L_0$, $\hat {\bar
  L}_0$ are Virasoro generators in the plane. The euclidean space-time
of the QFT forms a cylinder with length $\ell$ in the time direction, whose
classical free energy is ${\cal F}_\text{cyl} = -c (\pi/6)
(\ell/\delta) + {\cal F}_\text{ann.}$, with ${\cal F}_\text{ann.} =
-\ln \langle a | \exp(-2\pi (\ell/\delta) (\hat L_0 + \hat{\bar L}_0))
|b\rangle$ and boundary states $|a\rangle$, $|b\rangle$.  The first
term $\sim \ell$ is the extensive part of the cylinder energy, given
by the ground state of the QFT. If the lowest eigenvalue of $\hat L_0
+ \hat{\bar L}_0$ is zero (e.g. in unitary CFT's), for $\ell \gg
\delta\equiv 2\pi$ one has ${\cal F}_\text{ann.} \sim e^{-\eta
  \ell/2}$ where $\eta/2$ is the smallest positive eigenvalue of $\hat
L_0 + \hat{\bar L}_0$ that couples to $|a\rangle$ and $|b\rangle$
\footnote{Here it is assumed that the spectrum of $\hat L_0 +
  \hat{\bar L}_0$ is discrete which is not necessarily the case for $c
  \ge 1$.}. 
The decay of the two-point correlation function of the corresponding scaling 
field in unbounded space is also governed by the exponent $\eta$.
Since $\ell \to 2 \ln |\zeta|$ for large distance, we arrive at 
${\cal F}_\text{ann.} \sim |\zeta|^{-\eta}$ \cite{Eisenriegler:2004fk,Eisenriegler:2006uq}.

Next, consider
the QFT on an interval of length $\ell$ with the Hamiltonian $\hat H
= (\pi/\ell)( \hat L_0 - c/24 )$.  The cylinder is obtained as
the euclidean space-time of the QFT by choosing the time direction now
along the circumference $\delta$. For $\ell \ll \delta=2\pi$ this
yields ${\cal F}_\text{cyl}= - c (\pi/24) (\delta/\ell) - \ln
\text{Tr} e^{-\pi (\delta/\ell) \hat L_0}$. If the smallest eigenvalue
of $\hat L_0$ is $\tilde\eta/2$, for $\ell\to 0$ one has ${\cal
  F}_\text{ann.} \to \pi^2(\tilde \eta -c/12)/\ell$. Since $\ell$ is
given by the mutual capacitance, one has for smooth surfaces the short
distance expansion
\begin{equation}
  \label{eq:2.5}
  \frac{1}{\ell} = \sqrt{\frac{R_1R_2}{2(R_1+R_2)d}} +
  \frac{R_1^3+R_2^3}{12(R_1+R_2)^{5/2}}
\sqrt{\frac{d}{2R_1R_2}} + {\cal O} (d^{3/2})\,,
\end{equation}
where $R_\alpha$ are the local radii of curvature at the closest
points of the boundaries $\partial S_\alpha$ with separation $d$.

{\it Asymptotic limits of the geometric free energy} --- The geometric contribution
 ${\cal  F}_\text{geo.}$ is independent of the CFT, and solely related to the electrostatic potential through the map $\tilde w(z)=e^{w(z)}$. 
The large distance behavior of $F_\text{geo}$ can be obtained from
a multipole expansion with respect to origins $Z_\alpha$ inside $S_\alpha$, 
which yields the convergent bipolar series \cite{Nehari:1952tw}
\begin{equation}
  \label{eq:4}
  w(z) \!=\! \ln \frac{z-Z_1}{z-Z_2} + \!\sum_{m=1}^\infty \!\frac{1}{m}
\!\left[ \frac{\hat Q_{1,m}}{(z-Z_1)^m}  - \frac{\hat
    Q_{2,m}}{(z-Z_2)^m}  \right] \,,
\end{equation}
where coefficients $\hat Q_{\alpha,m}$ can be expressed in terms of the
electrostatic T-matrix elements of the objects and so-called
translation matrix elements that couple multipole moments (MM) on different
objects \cite{Rahi:2009uq}. This yields a distance $\zeta=Z_2-Z_1$
dependence of the form $\hat Q_{\alpha,m}=q_{\alpha,m,1}+
q_{\alpha,m,2}/\zeta + {\cal O}(\zeta^{-2})$. We expand the Schwarzian
derivative $\{\tilde w, z\}$ for large $z-Z_\alpha$ and move the
contour integration of Eq.~\eqref{eq:1} to the $y$-axis. With
$\text{Re}(Z_1)<0$, $\text{Re}(Z_2)>0$, this yields to leading order
at large distance
\begin{equation}
  \label{eq:6}
F_\text{geo} = - \frac{(\hat Q_{1,1}^2 + \hat Q_{1,2}) (\hat Q_{2,1}^2
  + \hat Q_{2,2})}{\zeta^5} + {\cal O}(\zeta^{-6}) \, .
\end{equation}
In most cases of interest the $\zeta^{-5}$ decay of $F_\text{geo}$ is
subdominant to $\zeta^{-(\eta+1)}$ coming from $F_\text{ann.}$. There
can, however, be exceptions \cite{Friedan:1984} with $\eta > 4$ where
the geometric force is dominant.

The short distance behavior of $F_\text{geo}$ is more complex.
On physical grounds we expect that the net Casimir force 
is dominated by points of closets approach.
In the so called proximity force approximation 
(PFA)~\cite{Derjaguin:1934dq,Parsegian:2005ly}, the force between
{\it smoothly varying surfaces} is obtained by integrating the pressure
for parallel plates, evaluated at local separations. 
This procedure is indeed consistent with the short-distance contribution from 
$F_\text{ann.}\equiv -\partial_\zeta {\cal   F}_\text{ann.}$ that follows
from Eq.~\eqref{eq:2.5}.
However, there is no corresponding `parallel plate pressure' for the geometric
force, since the $\tilde w(z)$ is now a global conformal map with $\{\tilde w, z\}=0$.
(For the same reason $F_\text{geo}=0$ between two circles.)
For PFA to remain valid, any contribution of $F_\text{geo}$ should be
sub-leading to $F_\text{ann.}$ as $d\to 0$, and we believe that
$F_\text{geo}$ approaches a shape-dependent constant in this limit.
PFA is not expected to hold for non-smooth surfaces, such as those with
sharp corners or tips. Indeed, for the case of needles (discussed below),
we find that both $F_\text{geo}$ and $F_\text{ann.}$ scale as $1/d$ for $d\to
0$. 

{\it Free energy of the annulus for specific models} --- The free energy for an
annulus is known exactly for certain CFTs.
For the free Gaussian field of a surface tension dominated interface (with
infinite capillary length), the free energy ${\cal F}_\text{ann.}$ on the annulus can be
expressed in terms of the Dedekind eta function
$\eta(\tau)=e^{i\pi\tau/12}\prod_{n=1}^\infty (1-e^{2\pi i n \tau})$
which is defined on the upper complex $\tau$-plane. One then obtains
\begin{align}
\label{eq:E_D_general}
{\cal F}_\text{ann.,D} &=
\frac{\pi}{6C}+
\frac{1}{2}\ln
\left(\frac{2\pi}{C}\right) 
+ \ln \eta\left(2i\over C\right) \,,\\
{\cal F}_\text{ann.,N} &=\frac{\pi}{6C} + \ln \eta\left(2i\over C\right) \,,
\label{eq:E_N_general}
 \end{align}
 for Dirichlet and Neumann conditions, respectively, and dropping
 an unimportant constant for the former.
For  small $C$ (large separations), this leads to 
${\cal F}_\text{ann.,D} \approx {\cal F}_\text{ann.,D,small C}  =(1/2)\ln (2\pi/C)$ 
and ${\cal F}_\text{ann.,N} \approx {\cal F}_\text{ann.,N,small C} =
 -e^{-4\pi/C}$.  This  distinct behavior at large separations
 follows from the absence of monopoles for Neumann boundary
 conditions. 
The Neumann result corresponds to $\eta=4$ and thus ${\cal
    F}_\text{ann.}$ scales the same way at large separations as ${\cal F}_\text{geo}$.
 For large
 $C$ (small separations) an expansion to {\it all orders} yields the  simple forms
\begin{align}
\label{eq:E_D_largeC}
{\cal F}_\text{ann.,D,large C}  & = \frac{\ln \pi}{2}
-\frac{\pi}{24}\,C  + \frac{\pi }{6 }\,\frac{1}{C} \,,\\
{\cal F}_\text{ann.,N,large C}  & = \frac{1}{2} \ln \frac{C}{2} 
-\frac{\pi}{24}\,C  + \frac{\pi }{6 }\,\frac{1}{C} \;.
\label{eq:E_N_largeC}
\end{align}
The accuracy of the approximations for small and large $C$ is remarkable,
with maximum errors of roughly $0.32\%$ and $1.6\%$ for Dirichlet and Neumann cases respectively.

For  $c=1/2$, CFT describes the continuum limit of the critical
Ising model. The free energy of the annulus depends on the boundary
conditions. For fixed spins on the boundaries, one has 
\begin{equation}
  \label{eq:7}
\nonumber
  {\cal F}_{\text{ann.},\pm} = \frac{\pi}{12C} - \ln \left[
    \chi_0\!\left(2i\over C\right) +
    \chi_{1\over 2}\!\left(2i\over C\right)  \pm \sqrt{2}\chi_{1\over
      16} \!\left(2i\over C\right)\right]\,,
\end{equation}
with upper (lower) sign for like (unlike) boundary conditions and with
Virasoro characters $\chi_0(\tau) =
[~\sqrt{\theta_3(\tau)/\eta(\tau)}+\sqrt{\theta_4(\tau)/\eta(\tau)}~]/2$,
$\chi_{1 \over 2}(\tau) =
[~\sqrt{\theta_3(\tau)/\eta(\tau)}-\sqrt{\theta_4(\tau)/\eta(\tau)}~]/2$,
$\chi_{1\over 16}(\tau) ~= \sqrt{\theta_2(\tau)/(2\eta(\tau))}$, where
$\theta_j(\tau)\equiv \theta_j(0|\tau)$ are Jacobi theta functions~\cite{Francesco:1997kx}. At large distance $\ell$ one has ${\cal F}_{\text{ann.},\pm} 
\to \mp \sqrt{2} e^{-\ell/8}$, and for vanishing $\ell$ the limits
${\cal F}_{\text{ann.},+} \to -\pi^2/(24 \ell)$ and  ${\cal F}_{\text{ann.},-} \to 23\pi^2/(24 \ell)$.
Both are consistent with the predicted  asymptotic behaviors with $\eta=1/4$,
$\tilde \eta_+=0$, and $\tilde \eta_-=1$.

\begin{figure}[ht]
\includegraphics[width= 1.\columnwidth]{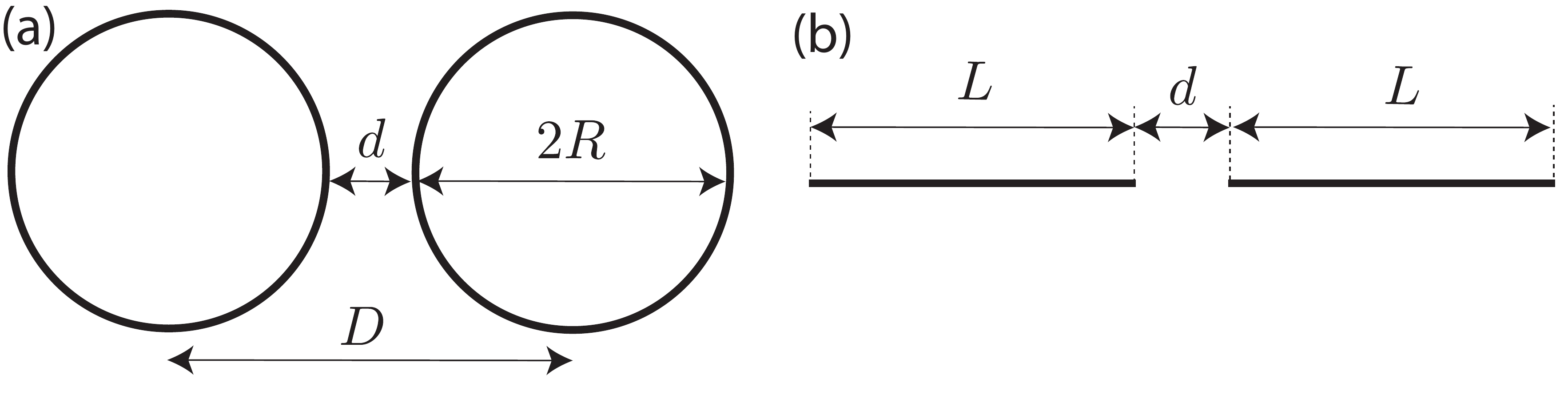}
\caption{Relevant length scales for (a) two circles, and (b) two aligned needles.}
\label{fig:geom}
\end{figure}

{\it Examples} --- We illustrate the power of our
general result with two examples for the free Gaussian field of
the interface (capillary wave) Hamiltonian for which $c=1$. 
The first case consists of two circles of equal radii $R$ and 
center-to-center separation $D$, as in Fig.~\ref{fig:geom}(a). 
This is the only (compact) geometry for which the geometric force 
$F_\text{geo}$ vanishes.  The mutual capacitance is 
$C= 2 \pi/\text{arccosh}[ \frac{1}{2} (D/R)^2 -1 ]$,~\cite{Smythe:1950vn}
and substitution into Eq.~\eqref{eq:E_D_general} yields at small surface-to-surface 
separation $d=D-2R \ll R$, the Dirichlet Casimir free energy
\begin{equation}
\label{eq:discs_s}
 {\cal F}_\text{D} =\frac{-\pi^2}{24 \sqrt{x}}\left[
   1+\!\left(\!\frac{1}{24}-\frac{4}{\pi^2}\!\right)\!x
+\!\left(\!\frac{1}{6\pi^2}-\frac{17}{5760}\!\right)\!x^2\!+\cdots\!\right],
\end{equation}
with $x=d/R$, where we have dropped a distance independent constant.  
At large distance one has  
\begin{equation}
\label{eq:discs_l}
 {\cal F}_\text{D}=\frac{1}{2} \ln \left (2\ln \frac{D}{R}\right) -
 \frac{1}{2 (D/R)^2 \ln D/R} +\ldots\,,
\end{equation}
which is in agreement with Ref.~\cite{Lehle:2006ys}.

\begin{figure}[h]
\includegraphics[width= 1.\columnwidth]{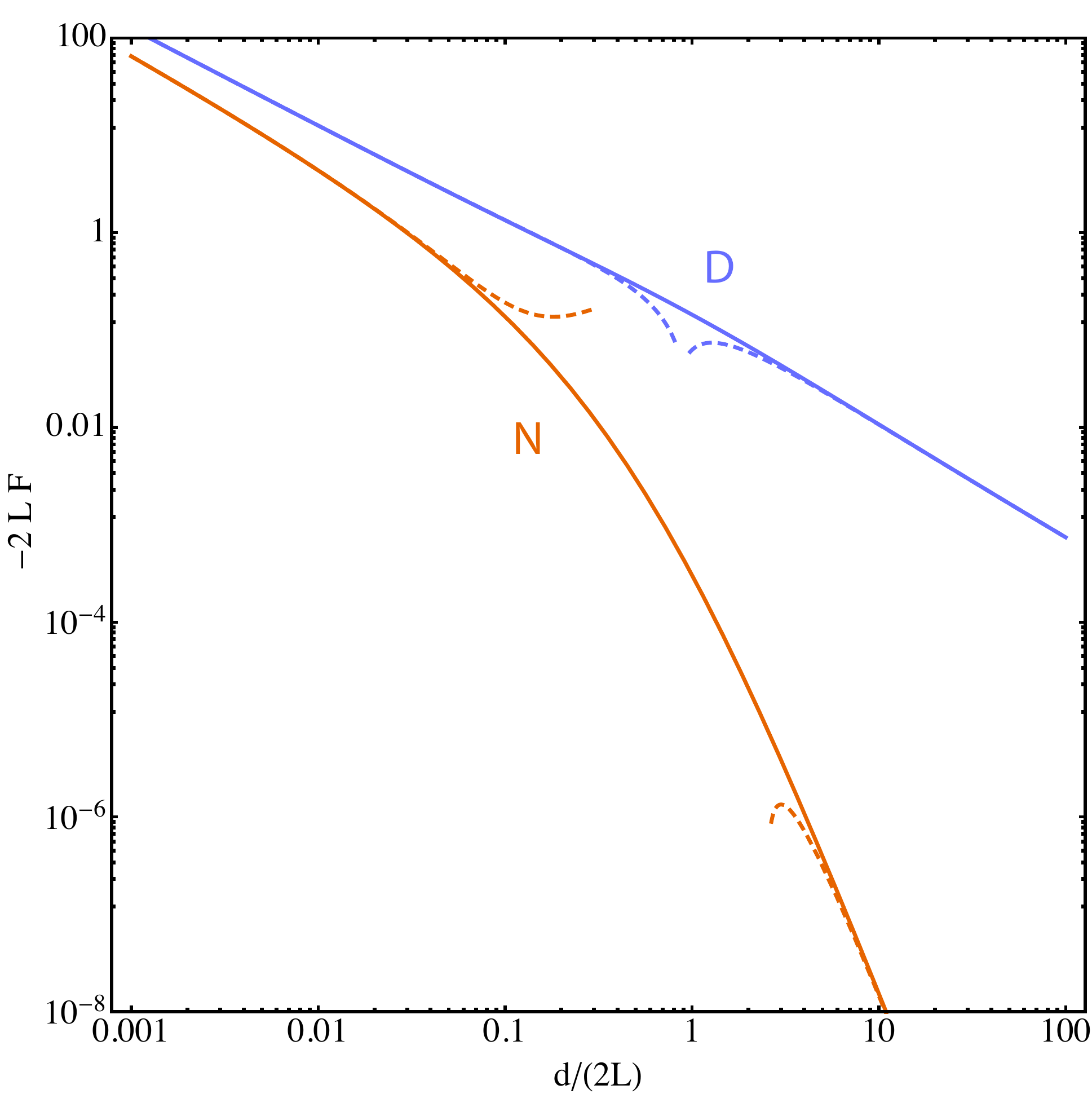}
\caption{The Casimir force $F$ between two aligned needles of length $L$
due to a scalar Gaussian field with Dirichlet (D) and  Neumann (N) boundary conditions, as a function of the tip-to-tip separation  $d$. 
Solid curves: exact result; dashed curves: short and large
distance expansions from Eqs.~\eqref{eq:aligned_needles_D_l}--\eqref{eq:aligned_needles_N_s}.}
\label{fig:force_plot}
\end{figure}

Next consider 
two aligned needles
of length $L$ and tip-to-tip distance $d$, as in Fig.~\ref{fig:geom}(b).
The conformal map $w(z)$ can be constructed by the Schwarz-Christoffel
transformation for polygons~\cite{Smythe:1950vn},
and the mutual capacitance is
$C=K(\sqrt{1-k^2})/K(k)$ where $K(k)$ is the complete elliptic
integral of the first kind, with $k=d/(2L+d)$. Contrary to 
smooth surfaces,  $F_\text{geo}$ does not go to a constant 
at short distances for needles which have singular curvature. 
In this limit, both ${\cal  F}_\text{ann.}$ and ${\cal  F}_\text{geo.}$  scale
logarithmically with separation for  D and N conditions.  At large
separation, the geometric component contributes to leading order only for N
conditions. The total Casimir force for Dirichlet conditions is given by
\begin{align}
  \label{eq:aligned_needles_D_l}
  2L\,F_\text{D} &= -\frac{1}{2x\ln(8x)} +\frac{1+\ln(8x)}{4x^2\ln^2(8x)} + {\cal O}\left(x^{-3}\right) \,,\\
\label{eq:aligned_needles_D_s}
 2L\, F_\text{D} &= -\frac{1}{8x} -\frac{1}{8}
+\frac{x}{4} +  {\cal O}\left(x^2\right)\,,
\end{align}
for large and small $x=d/(2L)$, respectively. For Neumann conditions the
two limits read
\begin{align}
  \label{eq:aligned_needles_N_l}
  2L\, F_\text{N} &= -\frac{1}{512\, x^5} +
  \frac{5}{1024\, x^6} + {\cal O}\left(x^{-7}\right) \\
\label{eq:aligned_needles_N_s}
2L\, F_\text{N} &= -\frac{1}{2x} \left( \frac{1}{4} -
  \frac{1}{\ln(4/x)}\right) -\frac{1}{8} \nonumber \\
& - \frac{1}{2\ln(4/x)} \left(
1+\frac{1}{\ln(4/x)}\right)
+ {\cal  O}\!\left(x\right) \, .
\end{align}
Figure~\ref{fig:force_plot} depicts the above asymptotic limits as dashed curves,
together with the exact result obtained from Eq.~\eqref{eq:1} with the
map $\tilde w(z)$ for two needles (solid curves).  
For D conditions the few terms of
Eqs.~\eqref{eq:aligned_needles_D_l},~\eqref{eq:aligned_needles_D_s}
give an accurate description at almost all separations.

The above examples nicely demonstrate how the exact form of the
Casimir force between two objects of arbitrary shape in a 2D CFT can
be obtained in terms of (i) the mutual capacitance $C$, (ii) the free
energy of the CFT on an annulus ${\cal F}_{\text{ann.}}$, and (iii) a
geometric contribution from the Schwarzian derivative of the map to
the annulus $\{\tilde w,z\}$.  $C$ an be easily computed with high
precision numerically; the asymptotic forms of ${\cal F}_{\text{ann.}}$ are known for all CFT. The geometric
  contribution to the force falls off as $\zeta^{-5}$ for large separations (for
  non-circular objects), its short distance behavior is non-trivially
  dependent on smoothness and other characteristics of the shape. To
  clarify this intricate shape dependence, calculations for other
  geometries are on the way. In particular, while not presented here
  for brevity, we have confirmed the $1/d$ divergence of $F_{\text{geo}}$
for finite wedges of arbitrary angle.

We thank R.~L.~Jaffe and B.~Duplantier for valuable discussions.  This
research was supported by the ESF Research Network CASIMIR, Labex PALM
AO 2013 grant CASIMIR, and the NSF through grant No. DMR-12-06323.

\end{document}